\documentclass[12pt]{article}
\usepackage{epsfig}

\def\beq{\begin{equation}}
\def\eeq{\end{equation}}
\input{tcilatex}
\begin{document}

\title{\textbf{Quark-Antiquark Potential and Generalized Borel Transform \thanks{%
Partially supported by CONICET and ANPCyT-Argentina.}}}
\author{L.N. Epele, H. Fanchiotti, C.A. Garc\'{\i}a Canal, M. Marucho \\
Laboratorio de F\'{\i}sica Te\'{o}rica\\
Departamento de F\'{\i}sica \\
Universidad Nacional de La Plata \\
C.C. 67 - 1900 La Plata \\
Argentina}
\date{}
\maketitle

\begin{abstract}
The heavy quark potential and particularly the one proposed by Richardson to
incorporate both asymptotic freedom and linear confinement is analyzed in
terms of a generalized Borel Transform recently proposed. We were able to
obtain, in the range of physical interest, an approximate analytical
expression for the potential in coordinate space valid even for intermediate
distances. The deviation between our approximate potential and the numerical
evaluation of the Richardson's one is much smaller than $\Lambda $ of QCD.
The $c \overline{c}$ and $b \overline{b}$ quarkonia energy levels agree
reasonably well with experimental data for $c$ and $b$ masses in good
agreement with the values obtained from experiments.
\end{abstract}

\vspace{5mm} \noindent Classification: 12.38 Aw - 12.38 Lg \newline
 \newpage

Among the different proposals for describing quark-antiquark interactions,
the Richardson's potential \cite{richardson}, due to its is simple
structure, has been the subject of continuous interest \cite{4241}. This
potential requieres, in the author words, the minimal number of parameters.
In fact, the only parameter entering the potential is the QCD related scale $%
\Lambda$. This potential, designed in order to present both asymptotic
freedom and linear quark confinement includes the single dressed gluon
exchange amplitude, namely 
\begin{equation}
V(q)= \frac{4}{3}\,\frac{\alpha_s(q^2)}{q^2}  \label{GEA}
\end{equation}

Asymptotic freedom is present as soon as one adopts for $\alpha _s(q^2)$ the
effective running coupling constant provided by the renormalization group.
Quark confinement is imposed by requiring that for $q$ small $V(q)$ behaves
as the inverse four power of $q$ that guarantees a linear behavior in $r$.
Then, the Richardson's potential reads 
\begin{equation}
V^R(q)=-\frac{16\,\pi ^2\,C_F}{\beta _0}\,\frac 1{q^2\,\ln (1+q^2/\Lambda
^2)}  \label{PR}
\end{equation}
where $C_F$ is a group coefficient.

It is clear that the explicit calculation of the QCD coupling constant in
position space is crucial when the Richardson's potential is to be applied
in a concrete case. This is because the Fourier transform of (\ref{PR})
provides the configuration space expression 
\begin{equation}
\overline{V}^R(r)=-\frac{C_F}r\,\frac{2\,\pi }{\beta _0}\,\overline{\alpha }%
(1/r)=-\frac{C_F}r\,\frac{2\,\pi }{\beta _0}\,\left[ a(1/r)-\Lambda
^2\,r^2\right]
\end{equation}
Here 
\begin{equation}
a(1/r)=1-4\,f(r)
\end{equation}
\begin{equation}
f(r)\equiv \int_1^\infty \frac{dq}q\,\frac{e^{-q\,\Lambda \,r}}{\left[ \ln
(q^2-1)\right] ^2+\pi ^2}  \label{1m4}
\end{equation}
This expression was only computed numerically. There exist some analytical
results corresponding to some asymptotic conditions. For example, for $%
\Lambda \,r\ll 1$, the Richardson's potential was shown to behave softer
than the Coulomb interaction \cite{limcoul}, namely 
\begin{equation}
\overline{V}^R(r)\rightarrow \frac 1{\Lambda \,r\,\ln (\Lambda \,r)}
\end{equation}
and for $\Lambda \,r\gg 1$ provides linear confinement.

Our main point in this paper is the calculation of the strong coupling
constant in position space starting from the last integral representation
eq.(\ref{1m4}). In so doing, we provide either an input for the Richardson's
potential in configuration space or to any other alternative proposal for
the quark potential including the original QCD running coupling constant or
any alternative expression \cite{rusos}. To this end, we fully analyze the
analytic structure of the integral in the Borel plane \cite{borel}. Then, we
are able to obtain the potential behaviour as a function of $r$, including
intermediate distances.

Any perturbative analysis starts from the general relationship between $V(q)$
and $\bar{V}(r)$ that ends with the corresponding relation between the
couplings $\alpha_s(q)$ and $\bar{\alpha}(1/r)$. Notice that in the
perturbative calculation, $\overline{\alpha}(1/r)$ coincides with $a (1/r)$.
Moreover, it has been shown \cite{HD} that one can write, for any static
potential, 
\begin{equation}
\overline{\alpha}(1/r) = \sum_n f_n\,\left[-\beta(\alpha_s)\,\frac{\partial}{%
\partial \alpha_s} \right]^n\,\alpha_s(q=\kappa/r)
\end{equation}
where $f_n$ are known constants, $\kappa = exp(\gamma_E)$ a constant and $%
\beta(\alpha_s) \equiv \mu^2\,\partial \alpha_s(q)/\partial \mu^2$. In the
case of the Richardson's potential, this series is asymptotically ($q^2 \gg
1 $) factorial divergent and its Borel sum does not exist, namely 
\[
\overline{\alpha}(1/r) \sim \alpha_s(q= \kappa/r)\,\sum_n
f_n\,\left[\beta_0\,\alpha_s(q=\kappa/r)\right]^n\,n! 
\]
Certainly, this expression provides sensible results only for very small
distances because at increasing distances the non-perturbative contributions
start to be important.

It is worth mentioning that the analytic behaviour of $\bar{\alpha}(1/r)$
has been studied \cite{PIRULO} by summing the divergent asymptotic series by
using the standard Borel formalism. Clearly, being the expression no Borel
summable, the Borel transform $B(s)$ has singularities for different values $%
s_k$ on the integration path of 
\[
\overline{\alpha}(1/r)= \int^{\infty}_0
exp\left[-s/\alpha(q=\kappa/r)\right]\,B(s)\,ds 
\]
Consequently, it can be defined only in principal value, showing ambiguities
coming from the exponential in this integral. This approach implies that the
non-perturbative contribution is considered of the same order of magnitude
as the ambiguity inherent to the method \cite{agli}. An additional problem
coming from the use of a perturbative $\alpha_s$ is the presence of the
Landau pole, conditioning the validity of any amplitude representing any
physical observable to a finite range of energy. In this respect, there is
an alternative proposal \cite{rusos} that starts from a modification of the $%
\alpha_s$ definition that avoids the Landau pole but retains the standard
properties. Nevertheless, this change implies a modification in the linear
confinement behaviour loosing the standard connection with the string
tension.

All the previous mentioned problems can be avoided by using the generalized
Borel Transform (GBT) that was introduced in Ref. \cite{marucho}. This
version of the Borel transform was originally defined on a finite lattice
but it can be readily adapted to the continuum, preserving all of its
characteristics.

The main vantage of this proposal comes from the fact that its analytic
properties have no ambiguities. Moreover, it allows to perform computations
in terms of a real and positive arbitrary parameter $\lambda$ avoiding the
implementation of perturbative expansions. The approach generally ends with
non-perturbative calculations of the saddle point type, when $\lambda$ takes
large arbitrary values. This is possible because the generalization implies
the definition of a valid Borel transformation for each value of the
parameter $\lambda$. Then, the better adapted value for each particular
problem can be chosen. In other words, when using the GBT, a function of $%
B_{\lambda}(s)$, a whole class of transformations is performed. It is found
that, as it should be, the results do not depend on $\lambda$. For the
particular case of the present paper, this can be summarize as

\[
a(1/r) = T^{-1}_{\lambda}\left[T_{\lambda}\left(a(1/r)\right)\right]
\,\,\,\,\,\,where\,\,\,\,\,\, T_{\lambda}\left(a(1/r)\right) \equiv
B_{\lambda}(s)\,\,\,\,\,\,\, for\,\,\,\,\, 0<r<\infty 
\]

We start by presenting our previous generalization \cite{marucho} of the
Borel transform of a function $f(r)$, namely 
\begin{equation}
B_{\lambda}(s) = \int_0^{\infty} exp[s/\eta(r)]\,\left[ \frac{1}{%
\lambda\,\eta(r)} + 1
\right]^{-\lambda\,s}\,f(r)\,d(1/\eta(r))\,\,\,\,\,;\,\,Re(s)< 0
\end{equation}
Among the properties of this definition we want to notice that it is valid
for any analytic function $\eta(r)$ in the interval $0<r<\infty$ that allows
to define 
\begin{equation}
u_{\lambda}(r) \equiv \frac{1}{\eta(r)} - \lambda\,\ln\left[\frac{1}{%
\lambda\,\eta(r)} + 1 \right]
\end{equation}
being monotonically increasing in the same interval if $\eta(r)$ is.
Consequently, the integral transform can be written as 
\begin{equation}
B_{\lambda}(s) = \int^{\infty}_0 exp[s\,u]\,f[r_{\lambda}(u)]\,\{1+ \lambda\,%
\eta[r_{\lambda}(u)]\}\,du\,\,\,;\,\,\,Re(s) < 0
\end{equation}
where $r_{\lambda}(u)$ is the inverse coming from the change of variables.
From the last expression it is clear that 
\begin{equation}
B_{\lambda}(s) = \int^{\infty}_0
exp(s\,u)\,L_{\lambda}[r(u)]\,du\,\,\,;\,\,\,Re(s) < 0
\end{equation}
is the Laplace transform of the function $L_{\lambda}[r(u)]$ implicitly
defined. That definition implies that $\eta(r)$ gives rise to an analytic
transformation in the negative Borel half-plane, such that its extension to
the other half-plane is also analytic with a cut on the real positive axis.
From this observation it is clear that $f(r)$ can be unambiguously expressed
in terms of the inverse Laplace transform integrated on the above mentioned
cut (for details see Ref. \cite{marucho}) 
\begin{equation}  \label{fr}
f(r) = \frac{1}{\lambda\,\eta(r) + 1}\,\int^{\infty}_0
exp[-s/\eta(r)]\,\left[\frac{1}{\lambda\,\eta(r)} + 1
\right]^{\lambda\,s}\,\triangle B_{\lambda}(s)\,ds
\end{equation}
As it was said before, the parameter $\lambda$ can take any real positive
non zero value generating a continuous family of transformations. A large
value of $\lambda$ could be useful because in this case asymptotic
techniques can be used in the calculations.

From this point on, a series of almost trivial calculations follows. Let us
only indicate the most important steps. Using the ansatz $1/\eta(r) =
\lambda\,\left[ exp(\Lambda\,r/\lambda) - 1 \right]$ which is well defined
for $0<r< \infty$, the function $u_{\lambda}(r)$ results 
\[
u_{\lambda}(r) = \lambda\,\left[ exp(\Lambda\,r/\lambda) - 1 \right] -
\Lambda\,r 
\]
Consequently, one can write 
\begin{equation}
B_{\lambda}(s) = \int_1^{\infty} dq\,H(q)\,\int_0^{\infty}
exp(s\,\lambda\,v)\,(v+1)^{-\lambda\,(s + q)} dv
\end{equation}
with 
\begin{equation}
H(q) = \frac 1{\left[ \ln \left( q^2-1\right) \right] ^2+\pi ^2}\,\frac1{q}
\end{equation}
where the change of variable $1/[\lambda\,\eta(r)] = v$ was introduced.
Notice that the last integral, for $Re(s) < 0$, represents the confluent
hypergeometric function \cite{bateman} $G(1,-\lambda\,s-\lambda\,q +2,
-s\,\lambda)$. Consequently, in this region $B_{\lambda}(s)$ is an analytic
function and when an analytic continuation to the positive half-plane of $s$
is performed, a cut appears. Introducing the discontinuity of the $G$
function, one gets 
\begin{equation}
\triangle B_{\lambda}(s) =
2\,\pi\,\lambda\,exp(-\lambda\,s)\,\int_1^{\infty} dq\,H(q)\,\frac{%
(\lambda\,s)^{\lambda\,(s+q) -1}}{\Gamma[\lambda(s+q)]}
\end{equation}

We can now transform back to obtain $f(r)$ from eq.(\ref{fr}) 
\begin{equation}  \label{di}
f\left( r\right) =\frac 1\pi\, \lambda ^2\,A_{\lambda} \left( r\right)
\int\limits_{-\infty }^\infty \int\limits_{-\infty }^\infty \exp \left[
G\left( w,t,r,\lambda \right) \right]\, dw\,dt
\end{equation}
where 
\[
A_{\lambda} \left( r\right) =1- \exp{\left[-\Lambda\,r/\lambda\right]} 
\]
and 
\begin{eqnarray}
G\left( w,t,r,\lambda \right) &=&-\lambda\, v\left( t\right)\, u_\lambda
\left( r\right) +t-\pi\, w-\ln \left[ w^2+1\right] -2\,\ln \left( q\left(
w\right) \right)  \nonumber \\
& & -\ln \left\{ \Gamma \left[ \lambda \left( \lambda\, v\left( t\right)
+q\left( w\right) \right) \right] \right\} +\lambda\, q\left( w\right)\, \ln
\left[ \lambda ^2\,v\left( t\right) \right] -\lambda ^2\,v\left( t\right) 
\nonumber \\
& & +\left( \lambda ^2\,v\left( t\right) -1\right)\, \ln \left[ \lambda
^2\,v\left( t\right) \right]
\end{eqnarray}
with 
\[
q(w) = \left[1 + e^{-\pi\,w}\right]^{1/2}\,\,\,\,\,\,;\,\,\,\,\,\,v(t) = e^t 
\]

The next step is to look for the asymptotic contribution in $\lambda$ of the
double integral in eq.(\ref{di}). To this end one can use the steepest
descent technique in the combined variables $(t,w)$. Consequently one first
computes the saddle points $t_0(r)$ and $w_o(r)$ and then one checks the
positivity condition \cite{jef}, in particular when the discriminant $%
D(t_0,w_0)$ of the second derivatives of $G$ at this point is positive. In
so doing one obtains 
\begin{equation}  \label{epf}
f\left( r\right) \simeq 2\,\lambda ^2\,A_\lambda \left( r\right)\, \exp
\left[ G\left( w_o\left( r\right) ,t_o\left( r\right) ,r,\lambda \right)
\right]\, \left[ \frac{\partial ^2G}{\partial w^2}\, \frac{ \partial ^2G}{%
\partial t^2}\ -\left( \frac{\partial ^2G}{ \partial w \partial t} \right)
^2 \right]^{-1/2}
\end{equation}
the saddle point being 
\begin{equation}  \label{wz}
t_o=\ln \left[ \frac{q_0\left( r\right) }{F\left( q_0\left( r\right) \right) 
}\right] =t_o\left( r\right)\,\,\,\,;\,\,\,\,\, w_o=-\frac 1\pi \ln \left[
q_0^2\left( r\right) -1\right] =w_o\left( r\right)
\end{equation}
where $q_0(r)$ is the solution of the implicit equation coming from the
extremes of the function $G$, namely 
\begin{equation}
r^2=\frac{F\left( q_0\right)}{\Lambda^2} \left[ F\left( q_0\right) +\frac
1{q_0}\right]
\end{equation}
with 
\begin{equation}
F\left( q_0\right) =\frac 2{q_0\,\left[ q_0^2-1\right] }\left[ 1-\frac{%
2\,q_0^2\ln \left( q_0^2-1\right) }{\left[ \ln \left( q_0^2-1\right) \right]
^2+\pi ^2 }\right]
\end{equation}

Notice that $F(q_0)$ should be positive and consequently $q_0 < 2.130156$.
On the other hand, from eq.(\ref{wz}), $q_0 >1$. In fact, moving $q_0$
between these values, the variable $r$ covers all the positive real axis in
a biunivocal way. Moreover, the condition $F(q_0) \neq 0$ implies that $r =
0 $ is excluded from the analysis. This is clearly not a drawback of the
method.

Going now to the expression (\ref{epf}) of $f(r)$, one finally finds, in the
saddle point approximation ($\lambda \rightarrow \infty $) 
\begin{equation}
f_{Ap}\left( r\right) =f\left( q_0\left( r\right) \right) \cong \frac{%
e^{-1/2}\,\sqrt{2}\,\sqrt{\left( F\left( q_0\right) +1/q_0\right) }}{2\,%
\sqrt{\pi \,D\left( q_0\right) }}\,\frac 1{\left[ q_0\right] ^{3/2}}\left[
q_0^2-1\right] \frac{\exp \left[ -q_0\,F\left( q_0\right) \right] }{\left(
\left[ \frac 1\pi \,\ln \left( q_0^2-1\right) \right] ^2+1\right) }
\label{final1}
\end{equation}

where

\begin{eqnarray}
D\left( q_0\right) &=&\frac{\pi ^2\left( q_0^2-1\right) }{2q_0^4}\left[ 1+%
\frac{q_0^3F\left( q_0\right) }2\right] +\frac{1-\left[ \frac 1\pi \ln
\left( q_0^2-1\right) \right] ^2}{\left[ \left[ \frac 1\pi \ln \left(
q_0^2-1\right) \right] ^2+1\right] ^2} \\
&&-\left[ \frac{\pi \left( q_0^2-1\right) F\left( q_0\right) }{2q_0}\right]
^2  \nonumber
\end{eqnarray}

Finally, the approximated expression for the potential is

\begin{equation}
V_{Ap}(r)=-\,\frac{2\,\pi C_F}{\beta _0}\,\left[ \frac{1-4\,f_{Ap}(r)}%
r-\Lambda ^2\,r\right]  \label{final}
\end{equation}

\medskip

In order to test the precision provided by the GBT, we have compared the
behavior in the coordinate space of our approximated analytical formula (24)
with the numerical integration of the exact expression (3). This comparison,
within the range of physical interest $0.1\,fm<r<1\,fm$ (see for example 
\cite{nora}, \cite{rusos}), has been performed for the Richardson's
potential free parameter $\Lambda =0.398\,GeV$ which he found to be the best
to fit the charmonium and botomonium lowest bound states. The corresponding
results are exhibited in Table I, being the relative error $5\%$ at most.

We have also evaluated, the charmonium and bottomoniun spectra following the
Richardson's algorithm. For this purpose, with the help of (24), we have
adjusted $m_c$ and $\Lambda$ to reproduce the best measured charmonium
energy levels $\psi(1S)$ and $\psi(2S)$ \cite{datos}. The resulting charm
quark mass is $m_c=1.363GeV$ and the scale size $\Lambda=0.434GeV$ (to be
compared with $m_c=1.491GeV$ and $\Lambda=0.398GeV$ obtained by Richardson).
In the bottomonium case we kept $\Lambda=0.434GeV$ assuming, like
Richardson, that the scale size should be a fundamental scale of the theory
independent from the quarkonium system. The bottom mass reproducing the best
measured bottomonium energy state $\Upsilon(2S)$ \cite{datos} results in $%
m_b=4780GeV$ . Notice that Richardson has fitted the botomonium fundamental
level $\Upsilon(1S)$ obtaining $m_b=4888GeV$. The results for both spectra
and the corresponding experimental values are presented in Tables II and III
respectively.

In Summary, we have obtained an analytic expression for the quark potential
valid for any value of $r$. Our results, in the leading order of the saddle
point approximation, give rise to absolute values of the uncertainties
always much smaller than $\Lambda $ of QCD ($\sim 0.4\,GeV$). It is worth
mentioning that any perturbation based calculation obtained by means of the
standard Borel transform, ends with uncertainties at least of the order of $%
\Lambda $ \cite{HD}. This result shows the ability and potentiality of the
previously introduced generalized Borel Transform.

As regards the quarkonia energy levels, our spectra prediction results
reasonably well when compared with the spectra obtained by Richardson. The
deviations of our results with respect to the corresponding experimental
values, are certainly smaller than the scale size $\Lambda$ (see Tables II
and III). Let us remark that our adjustment leads to quark masses, when
compared with those of Richardson, in best agreement with the values
reported in the particle data table \cite{datos}. The resulting $\Lambda $
parameter falls in the range of values expected from perturbative QCD.

There exist in the literature other simple potentials in the coordinate
space \cite{quigg} \cite{charmoniun} that can compete with our approximation
and that give similar results for the spectra. However these potentials are
strictly phenomenological, not having a direct connection with pertubatibe
QCD and in general, they have more free parameters to be adjusted.

Finally, our $Q\bar{Q}$ approximate potential, due to computing
simplicities, is clearly useful in any further calculations of interesting
physical quantities \cite{tt}.

\pagebreak

\noindent\textbf{Table I: Comparison between potentials for $\Lambda
=0.398GeV$}

\bigskip

\begin{tabular}{cccc}
$r$ $\left[ GeV\right] ^{-1}$ & $V_{Rich}\left[ GeV\right] $ & $V_{Ap}\left[
GeV\right] $ & $\left| \delta V\right| \bigskip $ \\ 
7.0 & 0.899941 & 0.899923 & 0.000018\medskip \\ 
5.0 & 0.554276 & 0.554244 & 0.000032\medskip \\ 
3.0 & 0.149355 & 0.149637 & 0.000281\medskip \\ 
1.0 & -0.571347 & -0.554343 & 0.017004\medskip \\ 
0.5 & -1.114402 & -1.062477 & 0.051926\medskip \\ 
0.2 & -2.208255 & -2.284687 & 0.086730
\end{tabular}

\smallskip \bigskip

\noindent\textbf{Table II: System }$c\overline{c}$

\bigskip $
\begin{array}{cccc}
l & E_{Exp}\left[ GeV\right] & E_{Rich}\left[ GeV\right] & E_{Ap}\left[
GeV\right] \bigskip \\ 
0 & 3.096 & 3.096 & 3.096\medskip \\ 
0 & 3.684 & 3.684 & 3.684\medskip \\ 
0 & 4.040\left( 4160\right) ^{*} & 4.096 & 4.127\medskip \\ 
0 & 4.415 & 4.440 & 4.506\medskip \\ 
1 & 3522 & 3.514 & 3.494\medskip \\ 
2 & 3770 & 3.799 & 3.786
\end{array}
$

\bigskip

\noindent Our parameters are $\left[ \Lambda=0.434GeV, m_c=1.363GeV\right] $
and the corresponding Richardson's ones $\left[ \Lambda =0.398GeV,
m_c=1.491GeV\right] $. \newline
$^{*}$ See \cite{faustov} for an alternative level assignment. \smallskip
\bigskip

\noindent\textbf{Table III: System }$b\overline{b}$

\bigskip $
\begin{array}{cccc}
l & E_{Exp}\left[ GeV\right] & E_{Rich}\left[ GeV\right] & E_{Ap}\left[
GeV\right] \bigskip \\ 
0 & 9.460 & 9.460 & 9.522\medskip \\ 
0 & 10.023 & 10.016 & 10.023\medskip \\ 
0 & 10.355 & 10.343 & 10.351\medskip \\ 
0 & 10.580 & 10.607 & 10.620\medskip \\ 
1 & 9.900 & 9.896 & 9.910\medskip \\ 
1 & 10.260 & 10.249 & 10.252
\end{array}
$ \bigskip

\noindent Our parameters used are $\left[ \Lambda =0.434GeV,
m_b=4.780GeV\right] $ and the corresponding Richardson'sq ones $\left[
\Lambda =0.398GeV, m_b=4.888GeV\right] $. \bigskip


\begin{thebibliography}{99}
\bibitem{richardson}  J.L. Richardson, Physics Letters \textbf{B82}
(1979)272.

\bibitem{4241}  M. Beneke, Physics Letters \textbf{B434} (1998) 115.

\bibitem{quigg}  C. Quigg, J.L. Rosner, Physics Report \textbf{56} (1979)
167.

\bibitem{limcoul}  M. Peter, Nuclear Physics \textbf{B501} (1997) 471.

\bibitem{rusos}  A.V. Nesterenko, Physical Review \textbf{D62} (2000) 094028
and hep-ph 0010257.

\bibitem{borel}  M. Beneke, Physics Report \textbf{317} (1999) 1.

\bibitem{HD}  M. Jezabek, M. Peter and Y. Sumino, Physics Letteres \textbf{%
B428} (1998) 352.

\bibitem{PIRULO}  M. Pindor, hep-th 9903151.

\bibitem{agli}  U. Aglietti, Z. Ligeti, Physics Letters \textbf{B364} (1995)
75; M. Beneke, V,M, Braun, Nuclear Physics \textbf{B426} (1994) 301.

\bibitem{marucho}  L.N. Epele, H. Fanchiotti, C.A. Garc\'{\i }a Canal, M.
Marucho, Nuclear Physics \textbf{B 583} (2000) 454.

\bibitem{bateman}  H. Bateman, Higher Transcendental Functions, Mc Graw
Hill, New York (1953), vol.1.

\bibitem{jef}  H. Jeffrey and B.S. Jeffrey, Methods of Mathematical Physics,
Cambridge (1966) page 187.

\bibitem{nora}  N. Brambilla, A. Vairo, hep-ph 9904330.

\bibitem{datos}  The European Physical Journal C \textbf{V15} (2000).

\bibitem{charmoniun}  E. Eichten, K. Gottfried, T. Kinoshita, K.D. Lane,
T.M. Yan, Physical Review \textbf{D21} (1980) 203.

\bibitem{tt}  M.Jezabek, J.H. K$\ddot{u}$hn, M. Peter, Y. Sumino, T.
Teubner, Physical Review \textbf{D58} (1998) 014006.

\bibitem{faustov}  R.N. Faustov, V.O. Galkin, A.V. Tatarintsev, A.S.
Vshivtsev, Int. J. Mod. Phys. \textbf{A15} (2000) 209.
\end{thebibliography}
\end{document}